\documentclass[runningheads]{llncs}
\usepackage[T1]{fontenc}
\usepackage{graphicx}
\usepackage{times}
\usepackage{latexsym}
\usepackage{booktabs}
\usepackage{array}
\usepackage{multicol, multirow}
\usepackage{biblatex}
\bibliography{references}
\newcolumntype{P}[1]{>{\centering\arraybackslash}m{#1}}
\usepackage{hyperref}
\begin{document}
\title{Wikipedia-based Datasets in Russian Information Retrieval Benchmark RusBEIR}
%
%
\author{Grigory Kovalev\inst{1} \and
Mikhail Tikhomirov\inst{1} \and
Pavel Mamaev\inst{1} \and
Olga Babina\inst{2} \and 
Natalia Loukachevitch\inst{1}}
\authorrunning{G. Kovalev, N. Loukachevitch et al.}
%
\institute{Lomonosov Moscow State University, Russia \and
           South Ural State University, Chelyabinsk, Russia}
\maketitle              
\begin{abstract}
In this paper, we present a novel series of Russian information retrieval datasets constructed from the ``Did you know…'' section of Russian Wikipedia. Our datasets support a range of retrieval tasks, including fact-checking, retrieval-augmented generation, and full-document retrieval, by leveraging interesting facts and their referenced Wikipedia articles annotated at the sentence level with graded relevance. We describe the methodology for dataset creation that enables the expansion of existing Russian Information Retrieval (IR) resources. Through extensive experiments, we extend the RusBEIR research \cite{kovalev2025building} by comparing lexical retrieval models, such as BM25, with state-of-the-art neural architectures fine-tuned for Russian, as well as multilingual models. Results of our experiments show that lexical methods tend to outperform neural models on full-document retrieval, while neural approaches better capture lexical semantics in shorter texts, such as in fact-checking or fine-grained retrieval. Using our newly created datasets, we also analyze the impact of document length on retrieval performance and demonstrate that combining retrieval with neural reranking consistently improves results. Our contribution expands the resources available for Russian information retrieval research and highlights the importance of accurate evaluation of retrieval models to achieve optimal performance. All datasets are publicly available at HuggingFace\footnote{https://huggingface.co/collections/msu-rcc-lair/rusbeir-datasets-6720fb076978ab6a77f4f64c}. To facilitate reproducibility and future research, we also release the full implementation on GitHub\footnote{https://github.com/kaengreg/rusBeIR}.

\keywords{Information retrieval  \and Wikipedia \and Neural models \and Lexical models}
\end{abstract}
\section{Introduction}
    The field of information retrieval (IR) has become a cornerstone of modern technology, enabling users to efficiently navigate the vast digital information landscape. Central to advancements in this field are high-quality datasets, which serve as benchmarks for evaluating retrieval systems, training machine learning models, and exploring contextual relationships within text. While English-language datasets such as BEIR \cite{thakur2021beir}, TREC \footnote{https://trec.nist.gov}, and others have been instrumental in driving progress in IR, resources for other languages - particularly those with complex linguistic structures and rich cultural contexts - remain underdeveloped. 
    The Russian language remains underrepresented in information retrieval (IR) research, with most existing evaluation datasets for Russian lacking in scale, diversity, and often relying on translated or multilingual sources. To address this gap, the creation of the RusBEIR benchmark \cite{kovalev2025building} includes efforts to gather as many native Russian IR datasets as possible, as well as to develop new ones to expand and enrich the available resources for Russian IR.

    Wikipedia, as a collaboratively edited, multilingual knowledge repository, offers a compelling foundation for constructing comprehensive information retrieval datasets. This idea is supported by popular datasets like Natural Questions (NQ) \cite{hardeniya2016natural} and MIRACL \cite{zhang2023miracl}, which are based on Wikipedia, along with several other examples. The Russian Wikipedia alone contains over 2 million articles, encompassing a broad range of topics, high-quality content, and structured metadata. However, its potential for IR research has been underutilized.

    \begin{figure}[h!]
        \centering
        \includegraphics[width=1\linewidth]{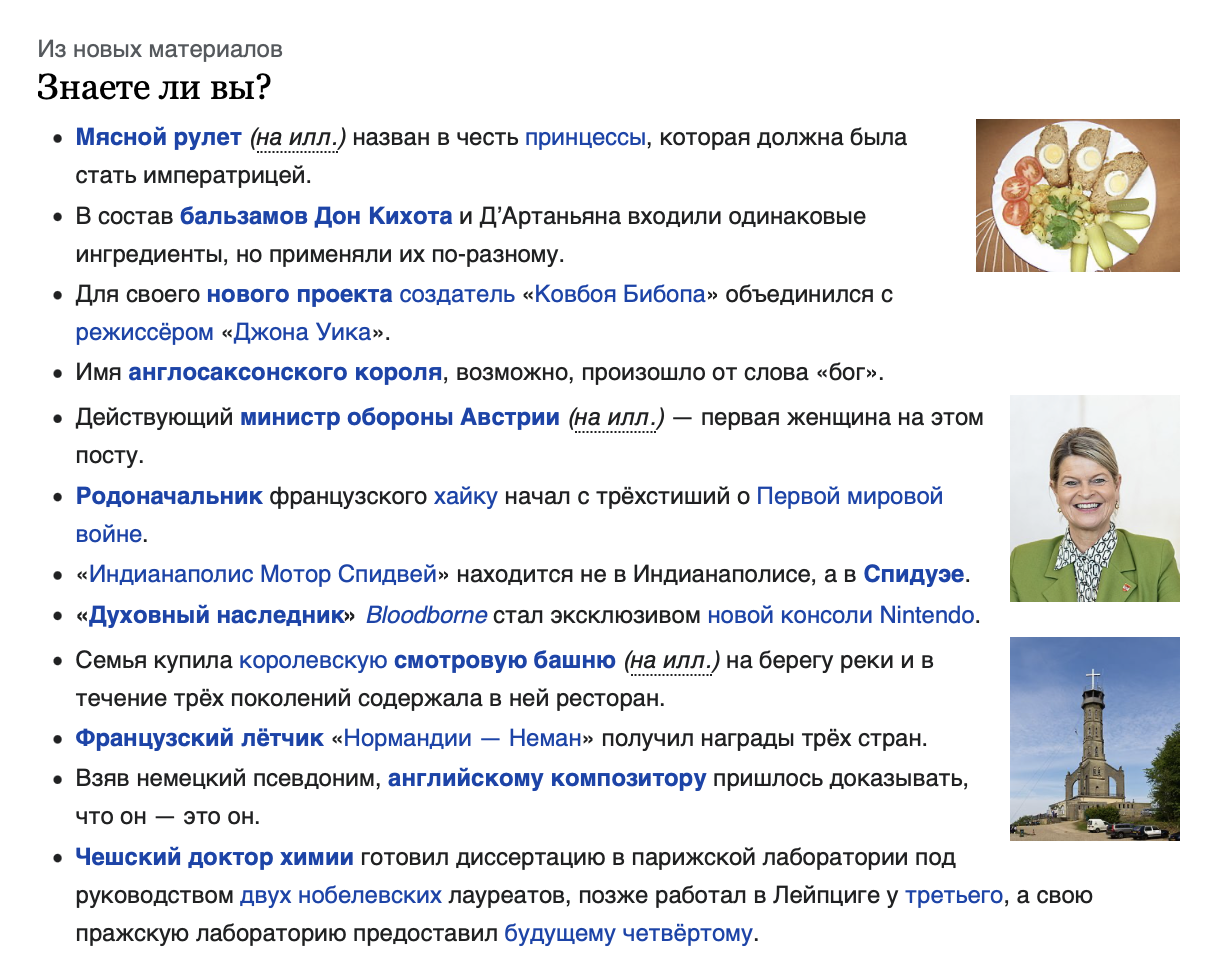}
        \caption{"Did you know..." section in Russian Wikipedia}
        \label{fig:did-u-know}
    \end{figure}
    
    In this paper, we suggest to utilize for information-retrieval evaluation the Wikipedia’s ``Did you know...'' section (fig. \ref{fig:did-u-know}). The section features interesting facts extracted from Wikipedia articles, with each fact including hyperlinks to the referenced articles. This section of Wikipedia, previously utilized primarily by scholars for academic research on the nature of information interest and data attractiveness \cite{prakash2015did,tsurel2017fun,kwon2020hierarchical}, has now been expanded to serve a broader audience. 
    
    Furthermore, the adaptable design of the corpus allows for the creation of diverse dataset sizes and configurations. For example, when the documents comprise complete articles, the dataset can function as a standard for full-document retrieval; conversely, if the corpus consists of individual sentences or their combinations, it is ideally suited for retrieval-augmented generation (RAG) tasks. A test sample of these datasets has already been incorporated into RusBEIR \cite{kovalev2025building}, an information-retrieval benchmark for Russian, but the datasets were not properly described. This paper examines the methodology behind the development of these datasets, explores their potential real-world applications, and extends the RusBEIR dataset collection while evaluating the impact of corpus size on model performance.

\section{Related work}
Throughout the time, Wikipedia has been a cornerstone for creating influential Natural Language Processing (NLP) datasets that drive progress in various language tasks.

    \textbf{SQuAD 2.0} \cite{rajpurkar2018know} is a reading comprehension dataset with over 150,000 questions, including 50,000+ unanswerable ones, based on Wikipedia articles. Queries are sentence-length questions; answers are text spans or ``unanswerable''. Crowdworkers created questions to test answer extraction and unanswerable question detection.
    
    \textbf{WikiQA} \cite{yang2015wikiqa} is a dataset for open-domain question answering with 3,047 Bing query log questions linked to Wikipedia pages. Contains 29,258 candidate answer sentences, 1,473 labeled correct. Queries are short phrases; documents are Wikipedia summary sentences. Data collected via query sampling and crowdsourced labeling.


\begin{table}[ht]
    \centering
    \resizebox{\linewidth}{!}{%
    \renewcommand{\arraystretch}{1} 
    \setlength{\extrarowheight}{0pt} 
    {\small
    \begin{tabular}{P{3cm}|P{2.5cm}|P{2.5cm}|P{3cm}|P{3.5cm}|P{1.5cm}}
        \toprule 
        \textbf{Dataset} & \textbf{Queries} & \textbf{Queries Source} & \textbf{Corpus} & \textbf{Data Collection Method} & \textbf{Relevance Level}\\
        \midrule
        SQuAD2.0 & Questions &  Generated by crowdworkers & Wikipedia articles & Crowdworkers verify questions & 2 \\
        \hline
        WikiQA & Search queries & Real user search queries & Wikipedia summary sentences & Bing query logs, crowdsourced labeling & 2 \\ \hline
        Natural Questions & Search queries & Real user search queries & Wikipedia pages & Google search queries, annotated answers & 2 \\ \hline
        WikiReading & Wikidata properties & Structured Wikidata facts & Wikipedia articles & Automated Wikidata-Wikipedia alignment & 2 \\ \hline
        DBpedia & Infobox properties & Structured Wikipedia infoboxes & Wikipedia infoboxes & Automated infobox parsing to RDF & - \\ \hline
        FEVER & Claims & Mutated Wikipedia sentences & Wikipedia intros & Sentence mutation, annotator verification & 2 \\ 
        \midrule
        RuBQ & Questions in Russian & Real user search queries & Wikipedia paragraphs & Search suggestions, annotated Wikidata links & 2 \\ \hline
        WikiOmnia & Generated questions in Russian & Auto-generated QA pairs & Wikipedia summaries & Fully automated generative pipeline & 2 \\ \hline
        SberQuAD & Questions in Russian & Generated by crowdworkers & Russian Wikipedia paragraphs & Crowdsourced annotation (Toloka) & 2 \\ 
        \midrule
        \multirow{2}{*}{\textbf{wikifacts-articles}} & & & \multirow{2}{*}{Full articles} & & \\
        & & & & & \\
        \cline{1-1} \cline{4-4}
        \multirow{2}{*}{\textbf{wikifacts-para}} & \multirow{4}{\linewidth}{\centering Assertions about facts} & \multirow{4}{\linewidth}{\centering Authored "Did you know" Wikipedia facts} & \multirow{2}{*}{Paragraphs} & \multirow{4}{\linewidth}{\centering Manual sentence-level annotation} & \multirow{4}{*}{3} \\
        & & & & & \\
        \cline{1-1} \cline{4-4}
        \multirow{2}{*}{\textbf{wikifacts-sents}} & & & \multirow{2}{*}{Sentences} & & \\
        & & & & & \\
        \cline{1-1} \cline{4-4}
        \multirow{2}{*}{\textbf{wikifacts-window}} & & & \multirow{2}{\linewidth}{\centering Sliding sentence chunks (2--6 sent.)} & & \\
        & & & & & \\
        \bottomrule
    \end{tabular}
    }}
    \vspace{5pt}
    \caption{Summary of Wikipedia-based datasets. The newly introduced Wikipedia Interesting Facts series are highlighted.}
    \label{tab:dataset_summary}
\end{table}
    
    \textbf{Natural Questions} (NQ) \cite{kwiatkowski2019natural} contains approximately 323,000 real Google Search queries paired with Wikipedia pages. Annotators mark long and short answers or note their absence. Queries are user searches; documents are Wikipedia passages. Data reflects real-world question answering challenges.

    \textbf{WikiReading} \cite{hewlett2016wikireading} is a dataset with 18M instances for predicting Wikidata values from Wikipedia articles. Queries are Wikidata properties (short phrases); documents are full articles. Data aligns Wikipedia text with Wikidata triples (entity, property, value).
    
    \textbf{DBpedia} \cite{auer2007dbpedia} is a structured knowledge base comprising over 100 million RDF triples from Wikipedia infoboxes, categories, and more. Queries are infobox properties (short phrases); documents are Wikipedia articles. Data extracted via automated parsing and ontology mapping.
    
    \textbf{FEVER} \cite{thorne2018fever} is a fact-checking dataset with 185,445 claims labeled SUPPORTED, REFUTED, or NOT ENOUGH INFO, verified against Wikipedia introductions. Queries are sentence-length claims; documents are introductory paragraphs. Claims created by modifying Wikipedia sentences.

    As mentioned above, Wikipedia serves as a valuable source of data that enables the creation of variety of datasets. This resource has also been utilized for the Russian language. Below are examples of datasets developed using Russian Wikipedia data:
    
    \textbf{RuBQ} \cite{rybin2021rubq} is a Russian knowledge base question answering (KBQA) dataset with 2,910 questions over Wikidata, including SPARQL queries. Queries are Russian questions from search suggestions; documents are Wikipedia paragraphs. Data collected via crowdsourcing and in-house annotation.
    
    \textbf{WikiOmnia} \cite{pisarevskaya2022wikiomnia} is a generative QA dataset comprising over 3.5 million verified question-answer pairs from Russian Wikipedia. Queries are auto-generated, sentence-length questions; documents are summary sections. Data created via automated pipeline using models like ruGPT-3 XL.
    
    \textbf{SberQuAD} \cite{efimov2020sberquad} is a Russian reading comprehension dataset with 50K+ paragraph-question-answer triples derived from Russian Wikipedia. Queries are sentence-length questions; answers
    are text spans without explicit start positions. Created following the SQuAD procedure, it shares a similar structure and task format but differs by having only one correct answer per question and no answer start indices. Data have been collected via crowdsourcing on Toloka.

    All these datasets demonstrate the remarkable adaptability of Wikipedia as a data source, demonstrating its capacity to support a wide range of natural language processing tasks - from reading comprehension and question answering to semantic retrieval and fact-checking. However, there are two main problems with existing Wikipedia-based datasets. First, most of these datasets are primarily available in English.
    Second, the method of creating such datasets often relies on crowdworkers generating queries, as previously mentioned. In this work, we introduce a new method for utilizing Wikipedia data to develop datasets for Russian.

\section{Wikipedia interesting facts}
    
We introduce a Wikipedia-based dataset, wikifacts-sents, derived from the ``Did you know...'' section (further ``interesting facts'') and the corresponding articles referenced in these facts. This dataset is designed to support fact-checking tasks and experiments with retrieval-augmented contexts, with the following key features:

\begin{itemize}
\item  Queries represent assertions about interesting facts, meaning the dataset can be categorized as fact-checking.
\item  Wikipedia facts are authored by Wikipedia contributors, 
eliminating the need for separate fact creation.
\item These facts are specially crafted by modifying initial assertions, often by combining information from multiple sources; this transformation can be described as a series of operations applied to the original Wikipedia content.
\item Each fact can have several supporting sentences: some sentences can contain full confirmation of the initial fact, while others may only confirm parts of the fact.
\item   Interesting facts are framed using descriptive mentions (such as ``sportsman'', ``city'', or ``book'') but include references to specific Wikipedia articles and titles, enabling identification of particular persons, objects, or events. This structure allows for evaluating LLMs’ knowledge of the facts and experimenting with retrieval-augmented contexts.    
\item The primary dataset is sentence-based, with facts confirmed at the sentence level, but its relevance markup enables fact verification in larger text units, such as paragraphs, articles, or text chunks of varying lengths, by aggregating relevant sentences. This flexibility allowed us to develop derived datasets with documents of different sizes.
\end{itemize}

    \begin{figure}
        \centering
        \includegraphics[width=1\linewidth]{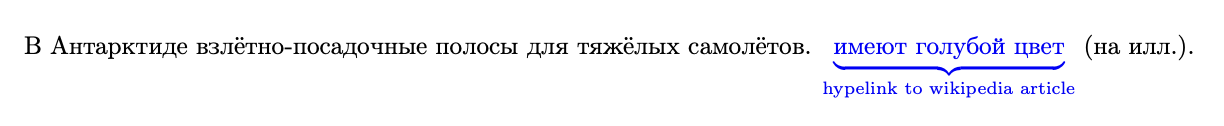}
        \caption{Examples of interesting facts from Russian Wikipedia}
        \label{fig:interesting-facts-exmp}
    \end{figure}

We conducted the annotation of sentences extracted from the Wikipedia articles, related to facts. In total, 55 university students were asked to identify relevant sentences. Each student was provided with a set of facts accompanied by referenced articles. The task was to read the articles associated with each fact and mark the sentences that appeared relevant, selecting from the predefined scoring options. A three-level relevance scoring system was employed: a score of 2 indicates that a sentence contains the complete information from the fact, a score of 1 shows that it includes only partial information, and a score of 0 denotes an irrelevant sentence.

A unique feature of our dataset is that crowdworkers are not required to create queries; their only task is to read the corresponding articles and mark relevant sentences. This is a much simpler task, making the work cheaper and allowing more data to be annotated for the same cost. Modern Large Language Models (LLMs) can also be used for the annotation process \cite{ma2024leveraging}. This can furthermore ease the process of creating similar datasets, making our dataset series even more actual and allowing to use the same methodology for the creation of similar wikipedia-based datasets in other languages.

At the end of the annotation process, we ended up with the wikifacts-sents dataset, containing 1,500,778 documents (sentences) and 5,433 queries, which serves as the foundation for creating multiple datasets derived from the ``Did you know...'' section.

\section{Wikipedia interesting facts series of datasets}

Based on the annotated wikifacts-sents data, additional seven wikifacts datasets have been produced. These datasets share a common set of queries but diverge in their corpus structures, providing a versatile platform for evaluation various retrieval approaches:

\begin{itemize} 
    \item \textbf{wikifacts-articles} dataset contains full articles referenced in the facts. Articles were systematically marked as relevant if at least one sentence within them was relevant to the corresponding fact. When multiple sentences showed different levels of relevance, the highest relevance score was chosen. This dataset features the longest documents across the series, making it an ideal resource for comprehensive full-document retrieval evaluation, particularly suited for testing the robustness of retrieval systems across extensive text corpora.
    \item \textbf{wikifacts-para} consists of paragraphs extracted from articles by splitting text at double newlines (``\textbackslash n\textbackslash n''). The documents in this variant are notably shorter than those in wikifacts-articles, yet they retain sufficient depth to serve as a valuable tool for evaluating full-text retrieval systems. This intermediate length allows for a balanced assessment of retrieval accuracy and efficiency.
    \item \textbf{wikifacts-window} is a specialized subset derived from the wikifacts-sents dataset, comprising sliding sentence chunks that range from 2 to 6 sentences. This configuration is specifically designed to evaluate Retrieval-Augmented Generation (RAG) models, which benefit from contextual text segments. Multiple versions of this dataset were generated with varying chunk sizes to assess the influence of text length on model performance. 
\end{itemize}

\begin{table*}[h!]
\centering
\renewcommand{\arraystretch}{1.25}
\resizebox{\linewidth}{!}{
\begin{tabular}{cc|cccc}
    \toprule
    \textbf{Task (↓)} & \textbf{Dataset (↓)} & \textbf{Relevancy} & \textbf{Dev}  & \textbf{Corpus} & \textbf{Avg. Word Lengths (D/Q)} \\ 
    \hline
    Information-Retrieval & wikifacts-articles & 3-level &  5,433 & 12,848  & 2059.1 / 11.4 \\ 
    Fact Checking &  wikifacts-para &  3-level &  5,433  & 176,364 & 150 / 11.4  \\ 
    Information-Retrieval & wikifacts-sents & 3-level & 5,433 & 1,500,778 & 17.6 / 11.4 \\ 
    Fact Checking &  wikifacts-window\_2 & 3-level & 5,433 & 1,500,776 & 35.3 / 11.4  \\ 
    Fact Checking &  wikifacts-window\_3 & 3-level & 5,433 & 1,500,774 & 52.9 / 11.4 \\ 
    Fact Checking &  wikifacts-window\_4 & 3-level & 5,433 & 1,500,772 & 70.5 / 11.4 \\ 
    Fact Checking &  wikifacts-window\_5 & 3-level & 5,433 & 1,500,770 & 88.1 / 11.4  \\ 
    Fact Checking &  wikifacts-window\_6 & 3-level & 5,433 & 1,500,768 & 105.8 / 11.4  \\ 
    \bottomrule
\end{tabular}
}
\vspace{5pt}
\caption{Overview of datasets.}
\label{tab:datasets}
\end{table*}

 To comply with the BEIR \cite{thakur2021beir} and RusBEIR\cite{kovalev2025building} formats, all datasets and their relevance annotations are provided as separate datasets on HuggingFace\footnote{https://huggingface.co/collections/msu-rcc-lair/rusbeir-datasets-6720fb076978ab6a77f4f64c}. The corpora and queries are stored in JSONL format, while the relevance judgments (qrels) are available in TSV files within datasets that have a qrels suffix matching the names of the corresponding corpus and queries.

It should be noted that the preliminary wikifacts data (540 facts, marked with a v0 suffix), constructed in the same manner, have already been included in the RusBEIR benchmark \cite{kovalev2025building}. Now we introduce a complete version  of the Wikipedia Interesting Facts dataset series, which currently comprises more than 5,000 facts.

To demonstrate that our datasets are not merely another collection of Wikipedia-based resources, but rather valuable benchmarks for information retrieval, we conduct an extensive series of experiments that highlight their practical utility.
Building on this, we further extend the RusBEIR benchmark \cite{kovalev2025building} by evaluating both previously used and newly developed models on these datasets. This broadens the range of assessed approaches and yields a more comprehensive leaderboard for Russian information retrieval.


\section{Models}

To thoroughly assess the effectiveness of information retrieval models using the Wikipedia Interesting Facts series of datasets, we employed traditional methods such as BM25, neural models based on bi-encoder and cross-encoder architectures, and a combination of traditional and neural approaches.

 \subsection{Preprocessing for BM25 model}
 The main baseline was established using the BM25 lexical model implemented in the Elasticsearch engine\footnote{https://www.elastic.co/}, with the language analyzer disabled to avoid stemming, which is known to be less suitable for the Russian language. We specially preprocessed data to be used as input for BM25.

 The text preprocessing method consists of the following steps:
    \begin{enumerate}
        \item \textbf{Lowercasing}: Converting all text to lowercase to ensure uniformity and eliminate case sensitivity.
        \item \textbf{Punctuation and Special Character Removal}: Using regex to remove non-alphanumeric characters, leaving only letters, digits, and spaces to reduce noise.
        \item \textbf{Space Normalization}: Removing extra spaces and trimming leading or trailing whitespace.
        \item \textbf{Tokenization}: Splitting text into individual words for processing.
        \item \textbf{Lemmatization}: Using the PyMorphy3 Russian morphological tool \cite{korobov2015morphological} to convert words into their dictionary forms, reducing data dimensionality while preserving word meaning. This approach is particularly effective for the Russian language due to its rich morphology, as it avoids the inaccuracies that stemming introduces by truncating words without context.
        \item \textbf{Stop Word Removal}: excluding overly frequent words that contribute little to the text content using the default stopword list provided by the NLTK package\footnote{https://www.nltk.org} \cite{hardeniya2016natural}, augmented with two Russian pronouns: ``which'' and ``such''.
    \end{enumerate}

\subsection{Neural baseline models}

In our experiments, we used neural models based on bi-encoder and cross-encoder architectures. Bi-encoders were employed as dense retrievers: they generate vector representations for both queries and documents, and their similarity is calculated using cosine similarity. Cross-encoders were used as rerankers, where each query is processed jointly with a document to estimate the likelihood of the document’s relevance to the query. Rerankers are typically applied to the top-k documents retrieved by either neural or lexical models to refine the results. In general, combined retrieval approaches tend to outperform individual models.

\noindent
The following pre-trained bi-encoders were used as dense retrievers:

    \textbf{LaBSE} bi-encoder \cite{feng2022language} was pre-trained with a translation ranking task. This allows to find sentence paraphrases in a single language or different languages.\footnote{https://huggingface.co/sentence-transformers/LaBSE}
  
    \textbf{Multilingual E5} in three variants: large\footnote{https://huggingface.co/intfloat/multilingual-e5-large}, base\footnote{https://huggingface.co/intfloat/multilingual-e5-base} and small\footnote{https://huggingface.co/intfloat/multilingual-e5-small} \cite{wang2024multilingual}. The multilingual E5 model was developed using a vast multilingual dataset, employing a weakly supervised contrastive learning approach with InfoNCE loss. It was further refined on high-quality, labeled multilingual datasets specifically for retrieval tasks. 
   
    \textbf{BGE-M3} model\footnote{https://huggingface.co/BAAI/bge-m3} \cite{chen2024bge} was pre-trained on a large multilingual and cross-lingual unsupervised data, and subsequently fine-tuned on multilingual retrieval datasets using a custom loss function based on the InfoNCE loss function.
    
    \textbf{USER-BGE-M3}\footnote{https://huggingface.co/deepvk/USER-bge-m3} is a sentence-transformer model designed to generate embeddings optimized for Russian. The model is initialized from the en-ru-BGE-M3 model\footnote{https://huggingface.co/TatonkaHF/bge-m3\_en\_ru},  a shrinked version of the BGE-M3 model, and then trained on the Russian datasets.
    
    \textbf{ru-en-RoSBERTa} \footnote{https://huggingface.co/ai-forever/ru-en-RoSBERTa} \cite{snegirev2024russian} is initialized from
    ruRoBERTa model \cite{zmitrovich2024family}\footnote{https://huggingface.co/ai-forever/ruRoberta-large} and then RoSBERTa embeddings were fine-tuned on Russian and English datasets. 
    
     \textbf{FRIDA} model\footnote{https://huggingface.co/ai-forever/FRIDA} is a large-scale, fine-tuned text embedding model that builds on the denoising approach originally introduced by T5, utilizing the encoder component of the FRED-T5 model\cite{zmitrovich2024family}. Trained on a bilingual Russian-English corpus, it received additional fine-tuning to enhance its capabilities in tasks such as information retrieval, classification, and paraphrasing, with a focus on the Russian language.

\begin{table}[h!]
    \centering
    \begin{tabular}{llcccc}
    \hline
    \textbf{Model} & \textbf{Based on} & \textbf{Parameters} & \textbf{Dim} & \textbf{Max input} \\ 
    \toprule
    Multilingual-E5-small & Multilingual-MiniLM & 118M & 384 & 512 \\ 
    Multilingual-E5-base & XLM-RoBERTa-base & 278M & 768 & 512 \\
    Multilingual-E5-large & XLM-RoBERTa-large  & 560M & 1024 & 512 \\ 
    BGE-M3 & BGE-M3 & 568M & 1024 & 8192\\
    USER-BGE-M3 & BGE-M3 & 359M & 1024 & 8192 \\
    LaBSE & LaBSE & 471M & 768 & 256\\
    RoSBERTa & SBERT & 404M & 1024 & 512 \\
    FRIDA & FRED-T5 & 823M & 1536 & 512\\
    \midrule
    bge-reranker-v2-m3 & BGE-M3 & 568M & 1024 & 8192 \\ 
    \bottomrule
    \end{tabular}
    \vspace{5pt}
    \caption{Model Specifications and Details}
    \label{tab:model_specs}
\end{table}

As shown in Table \ref{tab:model_specs}, different models support different maximum sequence lengths, which define their context windows. In our approach, we load both the model and tokenizer from HuggingFace, and set the \texttt{maxlen} parameter according to the model specification. Internally, the tokenizer produces at most \texttt{maxlen} tokens; if the input text is longer, it is truncated to this limit. As a result, the model processes only the first \texttt{maxlen} tokens, which are then used for subsequent computations.

Additionally, for state-of-the-art neural models such as mE5, BGE, and FRIDA, as well as the lexical BM25, we decided to use a reranker derived from one of these models to evaluate the effectiveness of such combinations. As demonstrated in \cite{zhang2024mgte}, the bge-reranker-v2-m3\footnote{https://huggingface.co/BAAI/bge-reranker-v2-m3} achieves strong performance compared to other rerankers. Consequently, we selected it as the reranking model for our study.

To enable a fair and robust comparison of retrieval models, we adopt the Normalized Discounted Cumulative Gain at rank 10 (NDCG@10) as the primary evaluation metric. NDCG@10 is widely regarded in information retrieval for its ability to assess not only the presence of relevant documents among the top retrieved results, but also their ranking quality


\section{Models' Performance on Wikipedia Interesting Facts}

The series of Wikipedia Interesting Facts datasets have been designed in a manner that enables a thorough assessment of model performance across diverse scenarios, including full-document retrieval, fact-checking, and precise fine-grained retrieval. The underlying concept is to evaluate models within the same domain while varying document size and structure, thereby uncovering valuable insights. 

Experiments with the selected models revealed important insights into the conditions under which neural models surpass lexical approaches, and vice versa. These findings offer a basis for understanding how model performance relates to specific document characteristics.

\begin{table*}[ht]
    \centering
    \resizebox{\linewidth}{!}{
    \begin{tabular}{lc|cccccccc|cccc}
        \toprule
        \multirow{2}{*}{Dataset (↓)} & \multicolumn{1}{c}{Lexical} & \multicolumn{8}{c}{Dense} & \multicolumn{4}{c}{Re-ranking} \\
        \cmidrule(lr){2-2} \cmidrule(lr){3-10} \cmidrule(lr){11-14}
        & BM25 & mE5-large & mE5-base & mE5-small & BGE-M3 & USER-BGE-M3 &  RoSBERTa & LaBSE & FRIDA & BM25+BGE & E5+BGE & BGE+BGE & FRIDA+BGE \\
        \midrule
        wikifacts-sents & 13.58 & 16.70 & 13.82 & 9.59 & 17.30 & 16.46 & 17.94 & 9.49 & \underline{20.79} & 16.31 & 17.79 & 17.65 & \textbf{22.10}\\
        wikifacts-para & \underline{44.57} & 28.81  & 28.48 & 17.51 & 37.70 & 40.29 & 36.60 & 11.27 & 41.34 & \textbf{51.13} & 39.77 & 46.89 & 48.82\\
        wikifacts-articles & \underline{74.95} & 60.62 & 56.21 & 51.69 & 64.11 & 66.02 & 59.80 & 27.25 & 60.54 & \textbf{74.97} & 70.39 & 71.22 & 69.31 \\
        \midrule 
        Avg & \underline{44.37} & 35.38 & 32.84 & 26.26 & 39.70 & 40.92 & 38.11 & 16.00 & 40.89 & \textbf{47.47} & 42.65 & 45.25 & 46.74 \\
        \bottomrule
    \end{tabular}
    }
    \vspace{5pt}
    \caption{Performance comparison across different models. The best results for each dataset are in bold; the results of the best single models are underlined.}
    \label{tab:results-wikifacts} 
\end{table*}

 Analyzing the results of lexical and neural models on the Wikipedia Interesting Facts datasets (presented in Table \ref{tab:results-wikifacts}), we observed that the lexical model BM25 performs increasingly well as document length increases. Notably, BM25 outperformed neural models on the "wikifacts-articles" and "wikifacts-para” datasets, which contain relatively long documents. This indicates a potential weakness of neural networks when processing longer documents: they may struggle to maintain consistent performance when compared to lexical approaches such as BM25.


\section{Impact of Document Length on the Model Performance} 
Neural models are constrained by limitations such as a maximum input sequence length, which can significantly impact their performance. Evaluations conducted using the extended set of the models from the RusBEIR benchmark \cite{kovalev2025building} revealed that neural models begin to encounter difficulties when processing longer documents, often failing to maintain consistent performance as document size increases.

To address this, we conducted an in-depth investigation into these capabilities, leveraging our newly developed series of wikifacts-window datasets. This dataset series is uniquely suited to provide a clearer understanding of the inherent limitations of neural models, particularly in handling texts of varying lengths. These findings offer critical insights that contribute to a more informed selection of models for specific tasks, ensuring optimal performance based on the nature of the documents involved.
\subsection{Lexical models vs Neural models}
 Drawing on the observations highlighted in Section 6, we conducted a detailed exploration of the relationship between text length and the performance of the lexical model BM25 using our wikifacts-window datasets. This collection of datasets serves as an ideal platform for such an analysis, as it encompasses documents of diverse lengths, ranging from short sentence clusters to larger text chunks, enabling a comprehensive assessment of retrieval performance across different scales.

\begin{table*}[ht]
    \centering
    \resizebox{\linewidth}{!}{
    \begin{tabular}{lc|cccccccc|cccc}
        \toprule
        \multirow{2}{*}{Dataset (↓)} & \multicolumn{1}{c}{Lexical} & \multicolumn{8}{c}{Dense} & \multicolumn{4}{c}{Re-ranking} \\
        \cmidrule(lr){2-2} \cmidrule(lr){3-10} \cmidrule(lr){11-14}
        & BM25 & mE5-large & mE5-base & mE5-small & BGE-M3 & USER-BGE-M3 &  RoSBERTa & LaBSE & FRIDA & BM25+BGE & E5+BGE & BGE+BGE & FRIDA+BGE \\
        \midrule
        wikifacts-sents & 13.58 & 16.70 & 13.82 & 9.59 & 17.30 & 16.46 & 17.94 & 9.49 & \underline{20.79} & 16.31 & 17.79 & 17.65 & \textbf{22.10}\\
        wikifacts-window\_2 & 21.27 & 24.98 & 20.98 & 15.70 & 25.56 & 24.90 & 25.03 & 12.38 & \underline{28.84} & 26.80 & 29.48 & 29.79 & \textbf{31.95} \\
        wikifacts-window\_3 & 24.94 & 27.30 & 23.97 & 18.65 & 27.81 & 27.97 & 26.40 & 12.00 & \underline{30.21} & 30.85 & 32.84 & 33.13 & \textbf{34.36}  \\
        wikifacts-window\_4 & 27.48 & 29.09 & 25.84 & 20.93 & 29.03 & 29.59 & 27.05 & 11.23 & \underline{30.71} & 33.37 & 34.76 & 34.44 & \textbf{35.38} \\
        wikifacts-window\_5 & 29.43 & 30.10 & 26.82 & 22.11 & 29.84 & 30.60 & 27.52 & 10.76 & \underline{31.22} & 35.32 & \textbf{36.27} & 35.48 & 36.05\\
         wikifacts-window\_6  & 31.54 & 31.42 & 28.11 & 23.34 & 31.15 & 32.08 & 28.39 & 10.88 & \underline{32.09} & 37.66 & \textbf{37.84} & 36.91 & 37.33 \\
        \bottomrule
    \end{tabular}
    }
    \vspace{5pt}
    \caption{Performance comparison across different models and Wikipedia Interesting Facts datasets. The best results for each dataset are in bold; the results of the best single models are underlined.}
    \label{tab:results-length} 
\end{table*}
    Table \ref{tab:results-length} shows that BM25 performs competitively with neural retrievers on longer texts, nearly matching their performance on the wikifacts-window\_6 dataset, where it is only 1.7 percentage points behind the best single model, FRIDA. Notably, BM25 outperforms mE5-large and BGE-M3 by 1.6 percentage points on this dataset, underscoring the robustness of lexical models in extended contexts. In contrast, on shorter datasets like wikifacts-window\_2 and wikifacts-window\_3, which have an average word count of 44, BM25 falls behind top neural retrievers by more than 15 percentage points, due to neural models’ superior ability to capture semantic nuances in brief texts. However, as document length exceeds 80 words, the performance gap narrows significantly, reflecting the input length limitations of most neural models. Notably, FRIDA and USER-BGE-M3, both non-multilingual models, consistently outperform other neural and lexical models across all the longest datasets. Their superior performance likely comes from fine-tuning on the target language, which allows for a deeper understanding that improves retrieval accuracy across both short and long texts. These results indicate that although lexical models like BM25 are efficient for longer documents, fine-tuned, language-specific neural models such as FRIDA and USER-BGE-M3 deliver consistently strong results.

\subsection{BGE max-length experiments}
As previously mentioned in Table \ref{tab:model_specs}, BGE is a unique model compared to others presented, as it has the ability to process up to 8192 tokens, allowing it to perform well on datasets with a high average word count. 
Therefore we chose to explore how the model's performance on the Wikipedia Interesting Facts datasets varies under different maximum length constraints. To mitigate domain-specific biases, we also incorporated datasets from RusBEIR that feature the longest texts across diverse domains.

Table \ref{tab:bge-comparison} shows the results of experiments conducted with different maximum-length parameters of the input.

\begin{table*}[ht]
    \centering
    \resizebox{\linewidth}{!}{
    \begin{tabular}{lcccccccccc}
        \toprule
          \multirow{2}{*}{Model (→)} & \multicolumn{10}{c}{Max-length / batch-size} \\
          \cmidrule(lr){2-11} 
          & \multicolumn{2}{c}{512 / 64} & \multicolumn{2}{c}{1024 / 64} & \multicolumn{2}{c}{2048 / 64} & \multicolumn{2}{c}{4096 / 32} & \multicolumn{2}{c}{8192 / 8} \\
          \cmidrule(lr){1-1} \cmidrule(lr){2-3} \cmidrule(lr){4-5} \cmidrule(lr){6-7} \cmidrule(lr){8-9} \cmidrule(lr){10-11}
         Dataset (↓) & BGE-M3 & USER-BGE-M3 & BGE-M3 & USER-BGE-M3 & BGE-M3 & USER-BGE-M3 & BGE-M3 & USER-BGE-M3 &  BGE-M3 & USER-BGE-M3 \\
        \toprule
        wikifacts-articles & 61.33 & 63.63 & 65.29 & 63.42 & 64.11 & \textbf{66.02} & 62.49 & 65.15 & 59.72 & 63.47\\
        wikifacts-para & 37.0 & 39.56 & 37.73 & 40.29 & 37.7 & \textbf{40.29} & 37.68 & 40.29 & 37.66 & 40.28 \\
        wikifacts-sents & \textbf{17.32} & 16.39 & \textbf{17.32} & 16.4 & 17.3 & 16.46 & 17.3 & 16.38 & 17.26 & 16.33 \\
        wikifacts-window\_2 & \textbf{25.58} & 24.95 & \textbf{25.58} & 24.95 & 25.56 & 24.9 & \textbf{25.58} & 24.95 & 25.57 & 24.9\\
        wikifacts-window\_3 & 27.83 & 27.97 & 27.83 & \textbf{27.98} & 27.81 & 27.97 & 27.82 & \textbf{27.98} & 27.83 & 27.96 \\
        wikifacts-window\_4 & 29.04 & 29.57 & 29.03 & \textbf{29.59} & 29.03 & \textbf{29.59} & 29.03 & \textbf{29.59} & 29.02 & 29.57 \\
        wikifacts-window\_5 & 29.84 & 30.6 & 29.83 & 30.6 & 29.84 & 30.6 & 29.82 & 30.6 & 29.84 & \textbf{30.61}\\
        wikifacts-window\_6 & 31.15 & 32.08 & 31.15 & 32.08 & 31.15 & \textbf{32.09} & 31.15 & 32.08 & 31.15 & 32.08\\
        \midrule
        rus-NFCorpus & 30.69 & 30.33 & 30.83 & 30.26 & \textbf{30.86} & 30.28 & 30.78 & 30.21 & 30.82 & 30.28\\
        rus-SciFact & 61.84 & 58.33 & 62.32 & 58.21 & \textbf{62.42} & 58.25 & 62.33 & 58.25 & 62.35 & 58.25\\
        rus-ArguAna & 50.68 & 46.65	& 50.71	& 46.58 & \textbf{50.75} & 46.52 & 50.72 & 46.54 & 50.73 & 46.54 \\
        Ria-News & 83.01 & 83.34 & 83.00 & 83.36 & 82.99 & \textbf{83.52} & 83.01 & 83.37 & 83.02 & 83.36\\
        \bottomrule
    \end{tabular}
    }
    \vspace{5pt}
    \caption{Performance comparison across BGE-M3 models with different max-length}
    \label{tab:bge-comparison} 
\end{table*}

We observe a consistent improvement in quality as the max-length increases from 512 to 2048. In some cases, models with max-lengths set to 4096 or 8192 achieve slightly better performance than those with 2048. However, these gains are not substantial enough to justify the additional computational cost and complexity associated with processing longer sequences.

Based on results it was determined that the BGE setup with a max-length of 2048 is the optimal choice, striking a balance between quality and efficiency. It provides strong performance across datasets while maintaining manageable resource requirements, making it a practical and effective configuration.

\section{Comparison to RusBEIR results}
The test sample from the Wikipedia Interesting Facts datasets was incorporated into RusBEIR, and since that time, the corpus size has expanded tenfold, significantly increasing the complexity of information retrieval tasks for neural models. This comparison - evaluating the same task with a substantially larger corpus - provides valuable insights into the challenges and nuances of employing neural models for information retrieval tasks, particularly in terms of scalability and performance under varying data volumes.

\begin{table*}[ht]
    \centering
    \resizebox{\linewidth}{!}{
    \begin{tabular}{lc|cccccccc|cccc}
        \toprule
        \multirow{2}{*}{Dataset (↓)} & \multicolumn{1}{c}{Lexical} & \multicolumn{8}{c}{Dense} & \multicolumn{4}{c}{Re-ranking} \\
        \cmidrule(lr){2-2} \cmidrule(lr){3-10} \cmidrule(lr){11-14}
        & BM25 & mE5-large & mE5-base & mE5-small & BGE-M3 & USER-BGE-M3 &  RoSBERTa & LaBSE & FRIDA & BM25+BGE & E5+BGE & BGE+BGE & FRIDA+BGE \\
        \midrule
        wikifacts-articles\_v0 & \underline{84.28} & 66.09 & 63.04 & 67.86 & 74.50 & 79.41 & 74.13 & 45.17 & 75.47 & \textbf{85.25} & 83.06 & 83.91 & 83.07 \\
        wikifacts-para\_v0 & \underline{61.31} & 50.15 & 49.51 & 34.71 & 54.55 & 57.53 & 50.66 & 14.78 & 55.17 & \textbf{66.61} & 59.95 & 63.76 & 63.85 \\
        wikifacts-sents\_v0 & 33.64 & 35.90 & 30.75 & 22.57 & 37.59 & 34.90 & 40.59 & 25.79 & \underline{46.53} & 39.96 & 38.53 & 39.20 & \textbf{49.41}\\
        wikifacts-window\_2\_v0 & 38.52 & 40.98 & 37.54 & 28.10 & 42.71 & 41.13 & 42.01 & 24.36 & \underline{47.23} & 46.80 & 48.26 & 48.74 & \textbf{52.15} \\
        wikifacts-window\_3\_v0 & 41.93 & 44.84 & 42.38 & 32.44 & 44.80 & 44.34 & 42.02 & 22.17 & \underline{47.77} & 50.23 & 51.49 & 52.30 & \textbf{53.30} \\
        wikifacts-window\_4\_v0 & 45.41 & 47.07 & 44.66 & 35.56 & 46.47 & 46.65 & 41.95 & 21.08 & \underline{47.94} & 53.84 & 54.24 & \textbf{54.35} & 54.27 \\
        wikifacts-window\_5\_v0 & 48.72 & \underline{49.93} & 47.18 & 37.70 & 48.53 & 49.22 & 43.83 & 21.13 & 49.13 & \textbf{57.45} & 57.38 & 57.02 & 55.87\\
         wikifacts-window\_6\_v0 & \underline{51.88} & 51.87 & 49.26 & 39.96 & 49.69 & 50.93 & 44.28 & 20.97 & 49.58 & \textbf{59.92} & 59.58 & 58.63 & 57.26 \\
        \midrule
        wikifacts-articles & \underline{74.95} & 60.62 & 56.21 & 51.69 & 64.11 & 66.02 & 59.80 & 27.25 & 60.54 & \textbf{74.97} & 70.39 & 71.22 & 69.31 \\
        wikifacts-para & \underline{44.57} & 28.81 & 28.48 & 17.51 & 37.70 & 40.29 & 36.60 & 11.27 & 41.34 & \textbf{51.13} & 39.77 & 46.89 & 48.82 \\
        wikifacts-sents & 13.58 & 16.70 & 13.82 & 9.59 & 17.30 & 16.46 & 17.94 & 9.49 & \underline{20.79} & 16.31 & 17.79 & 17.65 & \textbf{22.10}\\
        wikifacts-window\_2 & 21.27 & 24.98 & 20.98 & 15.70 & 25.56 & 24.90 & 25.03 & 12.38 & \underline{28.84} & 26.80 & 29.48 & 29.79 & \textbf{31.95} \\
        wikifacts-window\_3 & 24.94 & 27.30 & 23.97 & 18.65 & 27.81 & 27.97 & 26.40 & 12.00 & \underline{30.21} & 30.85 & 32.84 & 33.13 & \textbf{34.36}  \\
        wikifacts-window\_4 & 27.48 & 29.09 & 25.84 & 20.93 & 29.03 & 29.59 & 27.05 & 11.23 & \underline{30.71} & 33.37 & 34.76 & 34.44 & \textbf{35.38} \\
        wikifacts-window\_5 & 29.43 & 30.10 & 26.82 & 22.11 & 29.84 & 30.60 & 27.52 & 10.76 & \underline{31.22} & 35.32 & \textbf{36.27} & 35.48 & 36.05\\
         wikifacts-window\_6 & 31.54 & 31.42 & 28.11 & 23.34 & 31.15 & 32.08 & 28.39 & 10.88 & \underline{32.09} & 37.66 & \textbf{37.84} & 36.91 & 37.33 \\
        \bottomrule
    \end{tabular}
3    }
    \vspace{5pt}
    \caption{Performance comparison across different models and datasets. The best results for each dataset are in bold; the results of the best single models are underlined.}
    \label{tab:vers-compare} 
\end{table*}

The results in Table \ref{tab:vers-compare} demonstrate similar trends when comparing the performances of lexical, neural, and re-ranking models on the test version of Wikipedia Interesting Facts datasets (marked by v0 suffix) and on the extended version of the datasets.

Analyzing the complete version of the Wikipedia Interesting Facts datasets we can note, that the substantial corpus expansion significantly impacts all models, with performance reductions across the board due to increased retrieval complexity. While BM25 maintains strong performance on longer textual data, such as wikifacts-articles (74.95) and wikifacts-para (44.57), its relative effectiveness diminishes on shorter, sliding-window datasets, losing dominance on datasets like wikifacts-window\_5 and wikifacts-\\-window\_6. It should be noted that on wikifacts-window\_6, BM25 still outperforms mE5-large and BGE-M3, although the performance difference notably decreases, highlighting increased competition from neural retrieval models within increased dataset size.

Fine-tuned neural models specifically trained on Russian language data, such as FRIDA and USER-BGE-M3, consistently outperform multilingual neural models including the mE5 series and BGE-M3 starting from wikifacts-window\_3. FRIDA achieves the highest single-model scores across nearly all datasets, securing the leading position in individual neural model comparisons.

The combination of lexical or neural models with neural re-ranking demonstrates substantial improvements over individual retrieval approaches. Initially, the BM25+BGE combination delivered state-of-the-art results, particularly on longer texts, effectively combining lexical recall with semantic precision. However, a clear shift is observed in the expanded version of datasets, where the leadership of BM25+BGE notably declines for longer sliding-window datasets.
Specifically, on wikifacts-window\_5 and wikifacts-window\_6, the combination mE5-large with the BGE-reranker emerges as the top-performing approach, outperforming the previously leading BM25+BGE. This shift indicates the enhanced ability of dense retrieval combined with neural re-ranking to handle increased complexity and lexical variability in larger datasets.

Overall, these findings underline the necessity of integrating powerful neural rerankers with appropriate retrieval models to optimize performance in large-scale information retrieval tasks.

\section{Current leadership in RusBEIR}

With the introduction of new challenging datasets into RusBEIR, along with newly developed models, we updated the benchmark leaderboard to examine potential shifts in model rankings.

\begin{table}
    \centering
    \resizebox{\linewidth}{!}{
    \begin{tabular}{lc|cccccccc|cccc}
        \toprule
        Model (→) & \multicolumn{1}{c}{Lexical} & \multicolumn{8}{c}{Dense} & \multicolumn{4}{c}{Re-ranking} \\
        \cmidrule(lr){1-1} \cmidrule(lr){2-2} \cmidrule(lr){3-10} \cmidrule(lr){11-14}
        Dataset (↓) & BM25 & mE5-large & mE5-base & mE5-small & BGE-M3 & USER-BGE-M3 &  RoSBERTa &  LaBSE & FRIDA & BM25+BGE & E5+BGE & BGE+BGE & FRIDA+BGE \\
        \midrule
        rus-NFCorpus & \underline{32.33} & 30.96 & 26.90 & 26.79 & 30.86 & 30.28 & 27.24 & 18.53 & 29.40 & \textbf{34.83} & 33.18 & 32.46 & 32.38 \\ 
        rus-ArguAna & 41.49 & 49.06 & 39.40 & 39.59 & \underline{50.75} & 46.52 & 49.38 & 25.52 & 41.90 & 52.91 & \textbf{54.01} & 53.87 & 52.24\\ 
        rus-SciFact & \underline{65.60} & 63.49 & 63.46 & 60.46 & 62.42 & 58.25 & 53.90 & 29.07 & 63.43 & 70.40 & \textbf{71.34} & 69.64 & 70.09\\
        rus-SCIDOCS & 13.99 & 13.47 & 12.09 & 10.60 & \underline{15.04} & 14.46 & 14.43 & 8.17 & 12.78 & 15.31 & 15.98 & \textbf{16.21} & 15.33\\
        rus-TREC-COVID & 62.47 & 76.38 & 74.45 & 73.95 & 62.66 & 55.07 & 68.43 & 23.04 & \underline{82.42} & 73.46 & 83.11 & 77.66 & \textbf{85.97} \\
        rus-FiQA & 22.60 & 34.71 & 30.44 & 25.74 & \underline{38.16} & 37.09 & 32.62 & 7.19 & 36.50 & 29.37 & 39.19 & \textbf{40.21} & 38.36 \\
        rus-Quora & 61.34 & 80.18 & 78.62 & 74.99 & \underline{80.28} & 79.87 & 68.73 & 72.39 & 74.54 & 71.01 & \textbf{81.85} & 81.59 & 80.38\\
        rus-CQADupstack & 24.49 & 32.08 & 28.43 & 27.31 & \underline{32.57} & 31.40 & 27.85 & 21.36 & 27.69 & 28.86 & 33.78 & \textbf{33.90} & 31.44 \\
        rus-Touche & \underline{30.59} & 25.88 & 22.88 & 23.48 & 28.06 & 24.36 & 24.11 & 8.46 & 30.44 & \textbf{32.72} & 29.46 & 31.43 & 32.21\\
        \midrule
        rus-MMARCO & 15.25 & \underline{34.04} & 30.27 & 29.07 & 29.51 & 27.92 & 20.16 & 9.06 & 33.55 & 24.12 & \textbf{36.95} & 34.52 & 36.33 \\
        rus-MIRACL & 25.13 & 66.99 & 61.41 & 58.52 & 70.50 & 67.23 & 53.11 & 15.70 & \underline{71.91} & 41.51 & 75.90 & \textbf{76.44} & 76.36 \\
        rus-XQuAD & 96.19 & \underline{97.33} & 95.84 & 95.66 & 95.97 & 95.63 & 93.90 & 69.77 & 93.49 & 98.85 & \textbf{98.97} & \textbf{98.97} & 98.24 \\
        rus-XQuAD-Sentences & 82.36 & \underline{88.84} & 86.37 & 85.41 & 86.91 & 85.42 & 83.20 & 75.33 & 86.89 & 89.93 & \textbf{92.08} & 91.69 & 91.60\\
        rus-TyDi QA & 35.80 & 59.41 & 55.91 & 55.23 & 58.34 & 57.86 & 52.06 & 28.05 & \underline{59.84} & 50.12 & \textbf{66.20} & 65.78 & 65.84\\
        \midrule
        SberQuad-retrieval & 68.19 & 67.11 & 65.13 & 61.03 & \underline{68.26} & 67.03 & 63.59 & 37.54 & 62.68 & \textbf{70.34} & 69.41 & 68.21 & 65.34\\
        ruSciBench-retrieval &  36.69 & 50.81 & 45.74 & 42.93 & \underline{55.85} & 53.58 & 44.89 & 17.93 & 52.92 & 49.93 & 65.33 & \textbf{69.05} & 66.44 \\
        ru-facts &  92.56 & 93.65 & 93.55 & 93.06 & \textbf{93.91} & 93.77 & 93.66 & 93.10 & 93.74 & 92.72 & 92.87 & 92.87 & 92.90\\
        RuBQ & 37.33 & \underline{74.11} & 69.63 & 68.60 & 71.26 & 70.00 & 66.81 & 30.59 & 73.12 & 56.90 & \textbf{77.03} & 76.00 & 75.71\\
        Ria-News & 64.63 & 80.67 & 70.24 & 70.00 & 82.99 & \underline{83.52} & 78.85 & 61.57 & 82.92 & 78.12 & 86.22 & \textbf{86.85} & 86.73 \\
        \midrule
        wikifacts-articles & \underline{74.95} & 60.62 & 56.21 & 51.69 & 64.11 & 66.02 & 59.80 & 27.25 & 60.54 & \textbf{74.97} & 70.39 & 71.22 & 69.31 \\
        wikifacts-para & \underline{44.57} & 28.81 & 28.48 & 17.51 & 37.70 & 40.29 & 36.60 & 11.27 & 41.34 & \textbf{51.13} & 39.77 & 46.89 & 48.82 \\
        wikifacts-sents & 13.58 & 16.70 & 13.82 & 9.59 & 17.30 & 16.46 & 17.94 & 9.49 & \underline{20.79} & 16.31 & 17.79 & 17.65 & \textbf{22.10}\\
        wikifacts-window\_2 & 21.27 & 24.98 & 20.98 & 15.70 & 25.56 & 24.90 & 25.03 & 12.38 & \underline{28.84} & 26.80 & 29.48 & 29.79 & \textbf{31.95} \\
        wikifacts-window\_3 & 24.94 & 27.30 & 23.97 & 18.65 & 27.81 & 27.97 & 26.40 & 12.00 & \underline{30.21} & 30.85 & 32.84 & 33.13 & \textbf{34.36}  \\
        wikifacts-window\_4 & 27.48 & 29.09 & 25.84 & 20.93 & 29.03 & 29.59 & 27.05 & 11.23 & \underline{30.71} & 33.37 & 34.76 & 34.44 & \textbf{35.38} \\
        wikifacts-window\_5 & 29.43 & 30.10 & 26.82 & 22.11 & 29.84 & 30.60 & 27.52 & 10.76 & \underline{31.22} & 35.32 & \textbf{36.27} & 35.48 & 36.05\\
         wikifacts-window\_6 & 31.54 & 31.42 & 28.11 & 23.34 & 31.15 & 32.08 & 28.39 & 10.88 & \underline{32.09} & 37.66 & \textbf{37.84} & 36.91 & 37.33 \\
        \midrule
        \textbf{Avg.} & 43.58 & 50.67 & 47.22 & 44.52 & 50.99 & 49.90 & 46.88 & 28.06 & \underline{51.33} & 50.66 & 55.63 & 55.66 & \textbf{55.89} \\
        \bottomrule
    \end{tabular}
    }
    \vspace{5pt}
    \caption{Performance comparison across different models and datasets. The best results for each dataset are in bold; the results of the best single models are underlined.}
    \label{tab:rusBEIR-results-new} 
\end{table}

From Table \ref{tab:rusBEIR-results-new}, we confirm that newly added FRIDA continues to lead overall, consistently demonstrating strong performance both as a standalone model and when combined with reranking strategies.

As mentioned previously in Section 8, BM25 provides superior results on datasets with longer texts. The combination of BM25 and the BGE reranker performs comparably to single dense retrievers such as mE5-large and BGE-M3, which rank as the second and third best single models on our leaderboard. This further underscores that BM25 remains a robust and competitive model for information retrieval, while neural models excel in the context of shorter documents due to their ability to capture lexical semantics and provide deeper understanding.

Hybrid approaches that combine dense retrievers with rerankers after the retrieval stage show superior search quality, improving standalone results by an average of 9 percentage points, making them the best choice in terms of quality.

\section{Conclusion}
In this paper, we introduced and thoroughly analyzed the Wikipedia Interesting Facts series of datasets, specifically designed for Russian Information Retrieval tasks. Utilizing Wikipedia's ``Did you know...'' section, we created diverse datasets tailored for various retrieval tasks, ranging from full-document retrieval to fact-checking and fine-grained retrieval. These new datasets exhibited unique flexibility through varying document structures and sizes, enabling a comprehensive evaluation of different retrieval approaches.

Our experiments provided insights into model performance, revealing that lexical models such as BM25 remain competitive and robust, especially in full-document retrieval scenarios. However, neural models demonstrated superior ability to capture meaning within shorter text units. Language-specific neural models fine-tuned for Russian, such as FRIDA and USER-BGE-M3, outperformed multilingual variants on datasets featuring longer texts, highlighting the importance of accurately selecting retrievers for specific tasks. Despite the superior performance of Russian-specific models on wikifacts variants with larger sliding windows, their overall performance, as demonstrated in RusBEIR and Table \ref{tab:rusBEIR-results-new}, remains below that of multilingual variants.

Moreover, combining retrieval methods with neural reranking significantly enhanced performance, particularly in larger datasets. While the BM25+BGE combination initially established baselines for longer texts, further experiments on expanded datasets indicated that combinations involving dense neural retrievers, such as mE5-large with the BGE-reranker, surpassed previous leading models. For shorter datasets, the combination of FRIDA with the BGE-reranker demonstrated superior performance, surpassing mE5-large+BGE and BGE-M3+BGE. This underscores the importance of integrating neural rerankers to improve retrieval effectiveness.

Our novel dataset creation strategy not only expands the RusBEIR collection - facilitating more precise evaluation of information retrieval models - but also lays the groundwork for developing multilingual variants of these datasets. Our experiments indicate that evaluating models within the same domain under various scenarios can highlight critical nuances and ease the selection of optimal retrieval approaches.

\section*{Acknowledgments}
The study was supported by the grant of the Russian Science Foundation No. 25-21-00206 \footnote{https://rscf.ru/project/25-21-00206/}.

The research was carried out using the MSU-270 supercomputer of Lomonosov Moscow State University.

%
%
%
%
\printbibliography

@inproceedings{
    thakur2021beir,
    title={{BEIR}: A Heterogeneous Benchmark for Zero-shot Evaluation of Information Retrieval Models},
    author={Nandan Thakur and Nils Reimers and Andreas R{\"u}ckl{\'e} and Abhishek Srivastava and Iryna Gurevych},
    booktitle={Thirty-fifth Conference on Neural Information Processing Systems Datasets and Benchmarks Track (Round 2)},
    year={2021},
    url={https://openreview.net/forum?id=wCu6T5xFjeJ}
}

@article{wang2024multilingual,
  title={Multilingual e5 text embeddings: A technical report},
  author={Wang, Liang and Yang, Nan and Huang, Xiaolong and Yang, Linjun and Majumder, Rangan and Wei, Furu},
  journal={arXiv preprint arXiv:2402.05672},
  year={2024}
}

@article{snegirev2024russian,
  title={The Russian-focused embedders' exploration: ruMTEB benchmark and Russian embedding model design},
  author={Snegirev, Artem and Tikhonova, Maria and Maksimova, Anna and Fenogenova, Alena and Abramov, Alexander},
  journal={CoRR},
  year={2024}
}

@inproceedings{feng2022language,
  title={Language-agnostic BERT Sentence Embedding},
  author={Feng, Fangxiaoyu and Yang, Yinfei and Cer, Daniel and Arivazhagan, Naveen and Wang, Wei},
  booktitle={Proceedings of the 60th Annual Meeting of the Association for Computational Linguistics (Volume 1: Long Papers)},
  pages={878--891},
  year={2022}
}

@inproceedings{korobov2015morphological,
  title={Morphological analyzer and generator for Russian and Ukrainian languages},
  author={Korobov, Mikhail},
  booktitle={Analysis of Images, Social Networks and Texts: 4th International Conference, AIST 2015, Yekaterinburg, Russia, April 9--11, 2015, Revised Selected Papers 4},
  pages={320--332},
  year={2015},
  organization={Springer}
}

@article{chen2024bge,
  title={Bge m3-embedding: Multi-lingual, multi-functionality, multi-granularity text embeddings through self-knowledge distillation},
  author={Chen, Jianlv and Xiao, Shitao and Zhang, Peitian and Luo, Kun and Lian, Defu and Liu, Zheng},
  journal={arXiv preprint arXiv:2402.03216},
  year={2024}
}

@inproceedings{zmitrovich2024family,
  title={A Family of Pretrained Transformer Language Models for Russian},
  author={Zmitrovich, Dmitry and Abramov, Aleksandr and Kalmykov, Andrey and Kadulin, Vitaly and Tikhonova, Maria and Taktasheva, Ekaterina and Astafurov, Danil and Baushenko, Mark and Snegirev, Artem and Shavrina, Tatiana and others},
  booktitle={Proceedings of the 2024 Joint International Conference on Computational Linguistics, Language Resources and Evaluation (LREC-COLING 2024)},
  pages={507--524},
  year={2024}
}

@book{hardeniya2016natural,
  title={Natural language processing: python and NLTK},
  author={Hardeniya, Nitin and Perkins, Jacob and Chopra, Deepti and Joshi, Nisheeth and Mathur, Iti},
  year={2016},
  publisher={Packt Publishing Ltd}
}

@article{zhang2023miracl,
  title={Miracl: A multilingual retrieval dataset covering 18 diverse languages},
  author={Zhang, Xinyu and Thakur, Nandan and Ogundepo, Odunayo and Kamalloo, Ehsan and Alfonso-Hermelo, David and Li, Xiaoguang and Liu, Qun and Rezagholizadeh, Mehdi and Lin, Jimmy},
  journal={Transactions of the Association for Computational Linguistics},
  volume={11},
  pages={1114--1131},
  year={2023},
  publisher={MIT Press One Broadway, 12th Floor, Cambridge, Massachusetts 02142, USA~…}
}

@inproceedings{efimov2020sberquad,
  title={Sberquad--russian reading comprehension dataset: Description and analysis},
  author={Efimov, Pavel and Chertok, Andrey and Boytsov, Leonid and Braslavski, Pavel},
  booktitle={Experimental IR Meets Multilinguality, Multimodality, and Interaction: 11th International Conference of the CLEF Association, CLEF 2020, Thessaloniki, Greece, September 22--25, 2020, Proceedings 11},
  pages={3--15},
  year={2020},
  organization={Springer}
}

@inproceedings{rybin2021rubq,
  title={RuBQ 2.0: an innovated Russian question answering dataset},
  author={Rybin, Ivan and Korablinov, Vladislav and Efimov, Pavel and Braslavski, Pavel},
  booktitle={The Semantic Web: 18th International Conference, ESWC 2021, Virtual Event, June 6--10, 2021, Proceedings 18},
  pages={532--547},
  year={2021},
  organization={Springer}
}

@article{kwiatkowski2019natural,
  title={Natural questions: a benchmark for question answering research},
  author={Kwiatkowski, Tom and Palomaki, Jennimaria and Redfield, Olivia and Collins, Michael and Parikh, Ankur and Alberti, Chris and Epstein, Danielle and Polosukhin, Illia and Devlin, Jacob and Lee, Kenton and others},
  journal={Transactions of the Association for Computational Linguistics},
  volume={7},
  pages={453--466},
  year={2019},
  publisher={MIT Press One Rogers Street, Cambridge, MA 02142-1209, USA journals-info~…}
}

@inproceedings{zhang2024mgte,
  title={mGTE: Generalized Long-Context Text Representation and Reranking Models for Multilingual Text Retrieval},
  author={Zhang, Xin and Zhang, Yanzhao and Long, Dingkun and Xie, Wen and Dai, Ziqi and Tang, Jialong and Lin, Huan and Yang, Baosong and Xie, Pengjun and Huang, Fei and others},
  booktitle={Proceedings of the 2024 Conference on Empirical Methods in Natural Language Processing: Industry Track},
  pages={1393--1412},
  year={2024}
}

@inproceedings{yang2015wikiqa,
  title={Wikiqa: A challenge dataset for open-domain question answering},
  author={Yang, Yi and Yih, Wen-tau and Meek, Christopher},
  booktitle={Proceedings of the 2015 conference on empirical methods in natural language processing},
  pages={2013--2018},
  year={2015}
}

@inproceedings{thorne2018fever,
  title={FEVER: a Large-scale Dataset for Fact Extraction and VERification},
  author={Thorne, James and Vlachos, Andreas and Christodoulopoulos, Christos and Mittal, Arpit},
  booktitle={Proceedings of the 2018 Conference of the North American Chapter of the Association for Computational Linguistics: Human Language Technologies, Volume 1 (Long Papers)},
  pages={809--819},
  year={2018}
}

@inproceedings{rajpurkar2018know,
  title={Know What You Don’t Know: Unanswerable Questions for SQuAD},
  author={Rajpurkar, Pranav and Jia, Robin and Liang, Percy},
  booktitle={Proceedings of the 56th Annual Meeting of the Association for Computational Linguistics (Volume 2: Short Papers)},
  year={2018},
  organization={Association for Computational Linguistics}
}

@inproceedings{auer2007dbpedia,
  title={Dbpedia: A nucleus for a web of open data},
  author={Auer, S{\"o}ren and Bizer, Christian and Kobilarov, Georgi and Lehmann, Jens and Cyganiak, Richard and Ives, Zachary},
  booktitle={international semantic web conference},
  pages={722--735},
  year={2007},
  organization={Springer}
}

@article{hewlett2016wikireading,
  title={Wikireading: A novel large-scale language understanding task over wikipedia},
  author={Hewlett, Daniel and Lacoste, Alexandre and Jones, Llion and Polosukhin, Illia and Fandrianto, Andrew and Han, Jay and Kelcey, Matthew and Berthelot, David},
  journal={arXiv preprint arXiv:1608.03542},
  year={2016}
}

@article{pisarevskaya2022wikiomnia,
  title={Wikiomnia: generative qa corpus on the whole russian wikipedia},
  author={Pisarevskaya, Dina and Shavrina, Tatiana},
  journal={arXiv preprint arXiv:2204.08009},
  year={2022}
}

@inproceedings{prakash2015did,
  title={Did you know? mining interesting trivia for entities from wikipedia},
  author={Prakash, Abhay and Chinnakotla, Manoj K and Patel, Dhaval and Garg, Puneet},
  booktitle={Proceedings of the 24th International Conference on Artificial Intelligence},
  pages={3164--3170},
  year={2015}
}

@inproceedings{tsurel2017fun,
  title={Fun facts: Automatic trivia fact extraction from wikipedia},
  author={Tsurel, David and Pelleg, Dan and Guy, Ido and Shahaf, Dafna},
  booktitle={Proceedings of the Tenth ACM International Conference on Web Search and Data Mining},
  pages={345--354},
  year={2017}
}

@inproceedings{kwon2020hierarchical,
  title={Hierarchical trivia fact extraction from Wikipedia articles},
  author={Kwon, Jingun and Kamigaito, Hidetaka and Song, Young-In and Okumura, Manabu},
  booktitle={Proceedings of the 28th International Conference on Computational Linguistics},
  pages={4825--4834},
  year={2020}
}

@inproceedings{kovalev2025building,
  title={Building Russian Benchmark for Evaluation of Information Retrieval Models},
  author={Kovalev, Grigory and Tikhomirov, Mikhail and Kozhevnikov, Evgeny and Kornilov, Max and Loukachevitch, Natalia},
  booktitle={Proceedings of the International Conference “Dialogue},
  volume={2025},
  year={2025}
}

@article{ma2024leveraging,
  title={Leveraging large language models for relevance judgments in legal case retrieval},
  author={Ma, Shengjie and Chu, Qi and Mao, Jiaxin and Jiang, Xuhui and Duan, Haozhe and Chen, Chong},
  journal={arXiv preprint arXiv:2403.18405},
  year={2024}
}

\end{document}